\documentclass[%
 reprint,
 amsmath,amssymb,
 aps,
]{revtex4-2}

\usepackage{graphicx}
\usepackage{dcolumn}
\usepackage{bm}
\usepackage{amsfonts}
\usepackage{amsmath}
\usepackage{float}
\usepackage{caption}
\usepackage{color}

\usepackage{graphicx,subcaption}

\begin{document}

\preprint{APS/123-QED}

\title{Discrete Euler-Bernoulli Beam Lattices with Beyond Nearest Connections}

\author{R.~G. Edge, E. Paul, K.~H. Madine, D.~J. Colquitt, T.~A. Starkey, G.~J. Chaplain}

\begin{abstract}

The propagation of elastic waves on discrete periodic Euler-Bernoulli mass-beam lattices is characterised by the competition between coupled translational and rotational degrees-of-freedom at the mass-beam junctions. We influence the dynamics of this system by coupling junctions with beyond-nearest-neighbour spatial connections, affording freedom over the locality of dispersion extrema in reciprocal space, facilitating the emergence of interesting dispersion relations. A generalised dispersion relation for an infinite monatomic mass-beam chain, with any integer order combination of non-local spatial connections, is presented. We demonstrate that competing power channels, between mass and rotational inertia, drive the position and existence of zero group velocity modes within the first Brillouin zone.
\end{abstract}

\maketitle


\section{\label{Sec: Introduction}Introduction}
%

The adaptability and functionality of lattices has perpetuated the long-standing investigation of periodic structures, with the fundamental principles and analysis traced back to the seminal work of Brillouin \cite{Brillouin1953}. Utilising discrete linear mass-spring toy models with single degrees-of-freedom has enabled the comprehensive study of a multitude of complex systems, with great utility in the study of scalar wave systems such as acoustics \cite{Chen2021}. There has been recent interest in discrete lattice systems with multiple degrees-of-freedom at the lattice sites \cite{Madine2021}, that are modelled not by masses and springs, but by masses and flexural beams. The analysis is founded on the simplistic yet powerful Euler-Bernoulli (EB) beam theory, with deformation pathways pertaining to both translational and rotational degrees-of-freedom at the lattice sites. The simplicity afforded by EB beam theory lends itself to a wide range of dynamical systems, facilitating a comprehensive range of studies \cite{Wang2013,Caddemi2012,Eringen}, including the coupling of flexural-torsional waves in elastic lattices at interfaces \cite{Madine2022}. 

Endeavours to ascertain enhanced control over wave dynamics has driven a vast amount of research, propagating concepts such as dispersion engineering \cite{Chaplainreconfig, Kazemi2023, Pu2010} and the development of highly unusual dispersive phenomena, due to the expanse of possible applications \cite{Smith2004, Craster2013}. Of note is the ability to engineer regions with very low, or zero-group-velocity (ZGV) in dispersion curves, that have implications for enhanced energy harvesting devices and vibration control \cite{Chaplainrainbowreflection,Chaplainrainbowtrapping}. The emergence of such extrema in dispersion curves has been attributed to a variety of mechanisms. Classically, ZGV modes almost always exist at the Brillouin Zone (BZ) boundaries (except at Dirac cones, for example), where the Bragg condition is satisfied and standing waves form. Symmetry breaking within the unit cell \cite{Nassar2020, Frenzel2017} has also been shown to render ZGV modes, lifting accidental degeneracies \cite{Chaplainrainbowreflection}. However, inspired by recent structures emulating `Roton-like' dispersion \cite{Chen2021,Martinez}, the inclusion of beyond-nearest-neighbour (BNN) couplings has re-emerged \cite{Brillouin1953} as a popular method to manipulate and even `draw' dispersion curves \cite{Kazemi2023}.

The exploration of passive, discrete lattice systems with multiple degrees-of-freedom at the lattice sites, such as the EB systems we consider here, affords more flexibility over the dispersive properties of a lattice due to the additional coupling between translational and rotational displacements of the masses, and the competing power channels that arise as a result; these competing power channels drive the extrema in the dispersion curves. 

In this paper, we expand on the discrete flexural systems considered in Refs. \cite{Madine2021, Madine2022}, and implement BNN connections in an infinite, 1D monatomic chain of masses coupled by massless EB beams, outlined in Section \ref{Sec: Theory}. The model, which we solve both analytically and numerically, encodes all the dynamics of the system in the nodal junctions, assuming the same material beam parameters between nearest neighbour (NN) and BNN. The result is a robust framework in the form of a generalised eigenvalue problem where the effective beam displacement is extracted from the eigenvectors; the displacement profile of the effective beam represents the form as if there were the BNN beams attached, which we discuss in the context of active control. In particular we encourage and expect the development of virtual BNN connections that emulate physical BNN connections, particularly where passive structures may become cumbersome and difficult to realise \cite{Chaplainreconfig,Kazemi2023}. 

Additionally, we investigate the dispersive properties of the lattice and explicitly calculate the energy flux associated with each deformation pathway. We then highlight the flexibility of the model, showing that the position in reciprocal space of ZGV modes within the first BZ can be tuned arbitrarily with the rotational inertia for a given order of BNN connections.

\section{\label{Sec: Theory}Theory}
\begin{figure}
    \centering
    \includegraphics[width=0.45\textwidth]{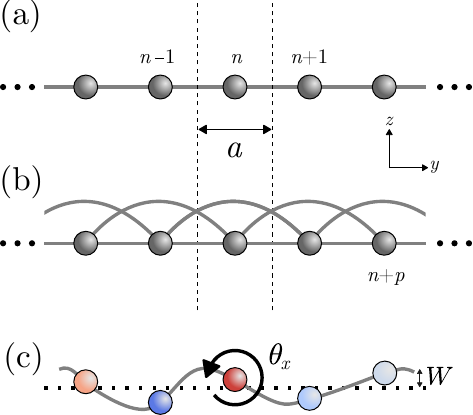}
    \caption{Schematics of (B)NN mass-beam systems. (a) Monatomic mass-beam chain with nearest-neighbours. (b) Monatomic mass-beam chain with $p = 2$ BNN. (c) Schematic of effective beam displacements, displaying the translation displacement at the nodal points $W$, and the rotation $\theta_x$, of the masses (highlighted by colourscale throughout) with anti-clockwise taken as positive (blue).}
    \label{fig:1D Lattice}
\end{figure}

This section briefly outlines the fundamental theory and provides an introduction to the basic mathematical procedure adopted from Ref. \cite{Madine2021}. We consider mass-beam lattices with any integer order of BNN connections that extend beyond the unit cell, without changing the spatial extent of the unit cell. The nodes in the lattices are treated as masses, of unitary separation ($a$=1), where the dynamic behaviour of the discrete system is encoded. Each node commands both mass and rotational inertia, and are connected to neighbouring nodes via beams that are governed by EB beam theory. The beams facilitate the coupling of translational and rotational nodal displacements through the shear force and flexural bending moment. A schematic of the system with and without BNN connections is shown in Fig.~\ref{fig:1D Lattice}.

Figure \ref{fig:1D Lattice}(a) depicts a lattice with local NN couplings, in which the connected beams have been assigned a positive or negative description depending on whether the beams connect between $n$ and $n\pm 1$ respectively. Additionally, Figure \ref{fig:1D Lattice}(b) portrays an example unit cell extended to include BNN connections of order $p=2$.
In accordance with EB beam theory, the out-of-plane flexural behaviour is governed by a fourth order equation \cite{Graff},  
\begin{equation}
    \frac{\rho A}{EI} \left(\frac{\partial ^2 W(y,t)}{\partial t^2}\right) +\frac{\partial ^4 W(y,t)}{\partial y^4}=0,
    \label{eq: Forth order}
\end{equation}
where $E$ is the Young's modulus, $I$ is the second moment of inertia (the product of which determines
the flexural rigidity). The density of the beams is $\rho$, $A$ is the cross-sectional area, and $W(y,t)$ is the time-dependent transverse displacement in the $z$-direction. In this paper we neglect the effects of beam buckling under deformation, thus assuming that the vertical displacement for the deformed beams is small compared to the beam lengths; the first-order spatial derivative of $W(y,t)$, corresponds to the angles of flexural rotation \cite{Jiao2019}. 

As the dynamics of the lattices are encoded within the nodes, the mass of each beam is assumed negligible compared to the mass contained within each nodal junction, thus we adopt a massless beam approximation $(\rho=0)$. We focus on time harmonic solutions and, without loss of generality, normalise \eqref{eq: Forth order} by the flexural rigidity such that the governing equation for the flexural motion of the beams are
\begin{align}
    \begin{split}
        \frac{\partial ^4 w(y)}{\partial y^4}=0.
        \label{eq: Governing forth order}
    \end{split}
\end{align}
Here $w(y)$ is the non-dimensionalised transverse displacement, the solutions of which are obtained using the following boundary conditions for beams lying parallel to $\hat{y}$ in the $xy$ plane. Below we detail boundary conditions at the positions of the masses as in Ref. \cite{Madine2021}, labelled by the index $n$, and include the order of the $p^{\text{th}}$ NN, with $n,p \in \mathbb{Z}$. Notably, careful attention is paid to the sign convention, ensuring appropriate coupling the between flexural behaviours \cite{Madine2021}, shown in Figure~\ref{fig:1D Lattice}, with $w$ referring to the transverse beam displacement, $w'$ being the first-order spatial derivatives. These boundary conditions are such that 
\begin{align}
    \begin{split}
        w(n-p)&=W_{-p}, \qquad w'_y(n-p)=\Theta_{-p}, \\ w(n)&=W_{0}, \qquad w'_y(n)=\Theta_{0}, \\ w(n+p)&=W_{p}, \qquad w'_y(n+p)=\Theta_{p}.
        \label{Eq: Boundary Conditions}
    \end{split}
\end{align}

Then, \eqref{eq: Governing forth order} returns solutions of the form
\begin{eqnarray}
    w(y)=C_0(y-n)^3+C_1(y-n)^2+C_2(y-n)+C_3,
\end{eqnarray}
where $C_0, C_1, C_2$, and $C_3$ are the coefficients obtained from the boundary conditions. Each coefficient is specific to the spatial order of the beams, where for a beam connecting the $n^{th}$ node to $(n\pm p)^{th}$ node the coefficients are
\begin{align}
    \begin{split}
        C_{0,\pm p} &= \pm \left(12\left|p^{-3}\right|(W_0-W_{\pm p})\right)+6\left|p^{-2}\right|\left(\Theta_{0}+\Theta_{\pm p} \right), \\ 
        C_{1,\pm p} &= 6\left|p^{-2}\right|\left(W_{\pm p}-W_0\right) \mp 2\left|p^{-1}\right|\left(2\Theta_{0}+\Theta_{\pm p} \right),\\
        C_{2,\pm p} &= \Theta_{0},\\
        C_{3,\pm p} &= W_0.        
        \label{Eq: Coefficients}
    \end{split}
\end{align}
In the following we denote the $p$ BNN connections as elements of the finite set $\mathcal{P} = \{1 \cup p\in\mathbb{N}^{+}\}$, so that nearest neighbours are always accounted for. Below we derive the dispersion relation of the systems by employing the Floquet-Bloch condition across the unit cell.

The wave behaviour of periodic systems is dependent on the coupling between neighbouring unit cells. Each beam connected at a nodal junction imposes some forcing and bending moment contribution to the dynamic behaviour of the lattice. The shear force, $\boldsymbol{F}_s(y)$, and flexural bending moment, $\boldsymbol{M}_{F}(y)$ supplied the beam to each mass are given by
\begin{eqnarray}
        \boldsymbol{F}_S(y)&=&-\frac{\partial ^3 w}{\partial y^3},\label{Eq: Shear Force}\\  \boldsymbol{M}_{F}(y)&=&\frac{\partial ^2 w}{\partial y^2},\label{Eq: Flex Moment} 
    \label{Eq: shear-torsional moments}
    \end{eqnarray} 
We introduce the generalised forcing vector on the $n^{\text{th}}$ node $\boldsymbol{F}_n=\left[F_{S}(y),\boldsymbol{M}_{F}(y)\right]^T$.
Thus can relate the force on the $n^{\text{th}}$ node from the $p^{\text{th}}$ neighbour applied from the positive and negative direction ($\pm$, as illustrated in Fig.~\ref{fig:1D Lattice}(a)) by:
\begin{align}
\begin{split}
    \boldsymbol{F}_{n,n+p} = A^+_p \boldsymbol{u}_n+B^+_p\boldsymbol{u}_{n+p} \\
    =
        \begin{bmatrix}
            -12p^{-3}& -6p^{-2}\\
            -6p^{-2} & -4p^{-1}
        \end{bmatrix}
        \begin{bmatrix}
            W_0\\
            \Theta_{0}
        \end{bmatrix}
        &+
        \begin{bmatrix}
            12p^{-3}& -6p^{-2}\\
            6p^{-2} & -2p^{-1}
        \end{bmatrix}
        \begin{bmatrix}
            W_p\\
            \Theta_{p}
        \end{bmatrix},\\\\
    \boldsymbol{F}_{n,n-p} = A^-_p \boldsymbol{u}_n+B^-_p\boldsymbol{u}_{n-p} \\
    =\begin{bmatrix}
            -12p^{-3}& 6p^{-2}\\
            6p^{-2} & -4p^{-1}
        \end{bmatrix}
        \begin{bmatrix}
            W_0\\
            \Theta_{0}
        \end{bmatrix}
        &+
        \begin{bmatrix}
            12p^{-3}& 6p^{-2}\\
            -6p^{-2} & -2p^{-1}
        \end{bmatrix}
        \begin{bmatrix}
            W_{-p}\\
            \Theta_{-p}
        \end{bmatrix}. 
    \end{split}
\end{align}
Here, $\boldsymbol{u}_n =\left[W_n,\Theta_{n}\right]^T$ is the generalised displacement vector, following a Hookean response. The integer $p$ is the order of the coupling regime of each beam, where $p=1$ is NN coupling, $p=2$ is next NN coupling, and so on. $A_p^{\pm}$ and $B_p^{\pm}$ are the stiffness matrices of the beams connecting the $n^{\text{th}}$ node to the $(n\pm p)^{\text{th}}$ node. Finally, the values of the terms contained within the stiffness matrices result from the solutions to \eqref{Eq: Shear Force} and \eqref{Eq: Flex Moment} expressed in matrix form. We obtain the equation of motion via application of Newton's second law, summing over the set of neighbours $\mathcal{P}$. We not that $p=1$ is always included to account for the NN response, whereby $\mathcal{P}=\{1\}$ returns the NN rigidity matrices detailed in \cite{Madine2021}. In this paper we focus our investigations on the $p=2$, and the $p=3$ cases for which $\mathcal{P}=\{1,2\}$, and $\mathcal{P}=\{1,3\}$ respectively. However this method can be applied to any number of $p\in \mathbb{N}^+$. This formalism lead to


\begin{align}
\begin{split}
    -M\omega ^2 \boldsymbol{u}_n &= \sum_{p\in \mathcal{P}}\ \big(A^+_p\boldsymbol{u}_n\\
    &+B^+_p\boldsymbol{u}_{n+p}+A^-_p\boldsymbol{u}_n +B^-_p\boldsymbol{u}_{n-p}\big).
\end{split}
\end{align}
Here, $M$ denotes the inertia matrix, 
\begin{equation}
    M=
    \begin{bmatrix}
            1& 0\\
            0 & \mu
        \end{bmatrix},
        \label{EQ: Mass Matrix}
\end{equation}
which describes the inertial properties of each node. $M$ has been normalised by the mass inertia $(\eta)$, such that the first component of $M$ is the normalised mass inertia associated with the translational displacements of the nodes, whilst $\mu=\frac{\widetilde{\mu}}{\eta}$ is the normalised rotational inertia associated with the nodes. As $\mu$ is the rotational inertia per unit nodal mass, under the thin beam approximation, we assume the rotational inertia ranges between $0<\mu<1$ and is such that $\mu<\eta$ \cite{Madine2021}.

The periodic nature of the monatomic lattice sanctions the use of Bloch's theorem, in which the behaviour observed within a single unit cell is used to infer the overall lattice response. Expression of the Bloch phase $(e^{ik\cdot p})$ translates the lattice behaviour between neighbouring unit cells, where $k$ represents the wave vector. The use of the discrete Fourier transform
\begin{align}
    \begin{split}
        \boldsymbol{U}=\sum e^{-ik \cdot {n}}\boldsymbol{u_{n}},
    \end{split}
\end{align}
yields the reciprocal space equation of motion for a 1D mass-beam lattice with any combination of BNN couplings:
\begin{equation}
    0 = \left[\omega ^2 M + \sum_{p\in\mathcal{P}}\ \left(A^+_p +B^+_pe^{ikp}+A^-_p +B^-_pe^{-ikp}\right)\right]\boldsymbol{U}.
    \label{Eq: Fourier Space EOM}
\end{equation}
Here $\boldsymbol{U}$ is the generalised Fourier displacement vector, and we denote the term in square brackets $G$, for use later. We determine the dispersion relation first analytically, using the solvability condition of \eqref{Eq: Fourier Space EOM} i.e. solving for the roots of the determinant
\begin{align}
    \begin{split}
        \sigma (\omega, k) &= 
        \mu\omega ^4 - \left(24\mu \alpha+\beta\right)\omega^2+24\alpha \beta+\gamma^2\\
        \alpha &= \sum_{p\in \mathcal{P}}\ p^{-3}\left(1-\cos\left(kp\right)\right)\\
        \beta &= \sum_{p\in \mathcal{P}}\ p^{-1}\left(8+4\cos\left(kp\right)\right)\\
        \gamma &= \sum_{p\in \mathcal{P}}\ 12ip^{-2}\sin\left(kp\right),
        \label{Eq: 1D Dispersion}
    \end{split}
\end{align}
and secondly numerically by recasting \eqref{Eq: Fourier Space EOM} as a generalised eigenvalue problem.

\section{\label{Sec: Results}Results and Discussion}
\begin{figure*}
    \centering
    \includegraphics[width=0.9\textwidth]{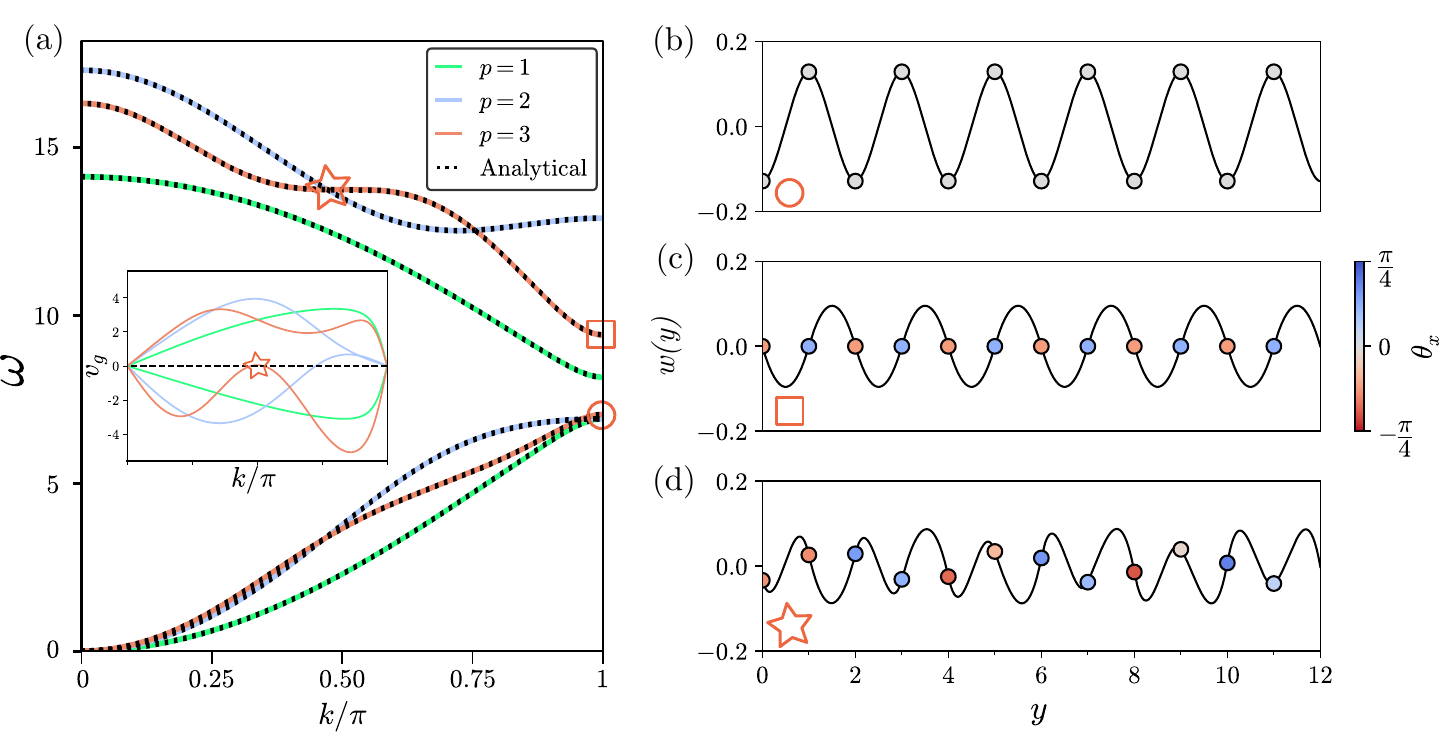}
    \caption{Dispersion curves and effective beam profiles, (a) Dispersion curves for 1D mass-beam chains with NN (green) and BNN ($p = 2$(3) in blue(red) respectively). Solid lines show numerical solutions via generalised eigenvalue problem with analytical solutions shown by dotted lines. The inset shows the group velocity, $v_g$. (b-d) Visualisation of the (effective) beam profile ($w(y)$) showing conventional standing wave behaviours (b-c) at the marked points in (a), for $p=3$ BNNs. Circles represent the nodal masses, colourbars illustrate the angular deformation. The character of the branches (i.e. translational vs rotational) is inferred from the standing wave solutions. In (d) we see an unusual standing wave due to the effective beam profile; there is no net power flow along the chain as we confirm below.}
    \label{Fig: 1D Dispersion}
\end{figure*}
 
In this section, we explore the dominant wave behaviours of discrete mass-beam lattices with different spatial couplings, and highlight the mechanism that drives the points of extrema in the dispersion curves within the first Brillouin Zone. We demonstrate that the simultaneous presence of both translational and rotational degrees-of-freedom at the nodes induces competing power channels within the lattice; it is these competing power channels that underpin this behaviour of elastic wave propagation in these lattices. 

We examine the analytical and generalised eigenvalue problem detailed in Section \ref{Sec: Theory}, for the first three orders of NN coupling. Figure~\ref{Fig: 1D Dispersion}(a) displays the dispersion relations, calculated using both methods, for $p$ = 1, 2 and 3, lattices with rotational inertia $\mu=0.06$.  Additionally, the inset shows the group velocity $v_g$ of both modes within the first Brillouin Zone, with the points of ZGV analytically described by,
\begin{align}
    \begin{split}
        v_g = 0 = \frac{\partial}{\partial k} \left[ \left(24\mu \alpha +\beta \pm \sqrt{\left(24\mu \alpha-\beta \right)^2-4\mu \gamma ^2 }\right)\right],
        \label{Eq: Vg=0}
    \end{split}
\end{align}
where the  $\pm$ denotes the upper and lower branches respectively.

In Fig. \ref{Fig: 1D Dispersion}(a) we observe the existence of two dispersion branches, as expected, due to the number of degrees-of-freedom: a transverse displacement in $\hat{z}$, and a rotation about $\theta_x$. The coupled nature of the translational and rotational behaviours prevents the sole attribution of a behaviour-type to a given branch. However, we label the primary behaviour for each mode based upon the behaviour at the first BZ boundary, where a standing wave results from either purely translational or rotational behaviours. 

The unitary period of the lattice dictates that the BZ boundary exists at $k=\pi$. By solving \eqref{Eq: 1D Dispersion} for $k=\pi$, it is possible to determine the form of both the upper and lower band edge solutions for arbitrary $p$. By doing so we obtain,
\begin{align}
\begin{split}
    \omega^2_{1} &= 48\left(1  +\sum_{p>1, \text{odd}} p^{-3}\right), \\
    \omega^2_{2} &= \frac{4}{\mu} \Big(1+ \sum_{p>1, \text{odd}} p^{-1}+\sum_{p>1, \text{even}} 3p^{-1} \Big),
    \label{}
\end{split}
\end{align}
for $p\in\mathcal{P}$ with $\omega^2_{1}$ being the squared solution of the lower branch at the band edge and $\omega^2_{2}$ is that of the upper branch at $\mu =0.06$. The effects of the NN lattice connections $p=1$ are included outside of the summations. Summations over $p$ are split into odd and even integer order of BNN; we see that \textit{only} the upper band edge frequency (the rotational standing wave) is affected by coupling to even order neighbours. 

The absence of a rotational dependence for $\omega^2_{1}$,  infers that the lower branch is primarily translational in nature at the BZ boundary. This is modelled in Figure \ref{Fig: 1D Dispersion}(b), where the flexural behaviour of the lower band edge solution is plotted for the $p=3$ case where the lattice links represent the effective displacement of the mass-beam system. Evidencing the translational behaviour via the nodal displacements. Conversely, as $\omega^2_{2}\propto \mu^{-1}$, this attests to a dependence on the rotational behaviour, inferring that the upper branches primary behaviour is rotational in nature at the BZ boundary. Similarly, this is illustrated in Figure~\ref{Fig: 1D Dispersion}(c). Given that the lattice dynamics are encoded within the nodes, the absence of a nodal displacement, verifies that the observed behaviour is primarily rotational in nature. 

\begin{figure}
    \centering
    \includegraphics[width=0.495\textwidth]{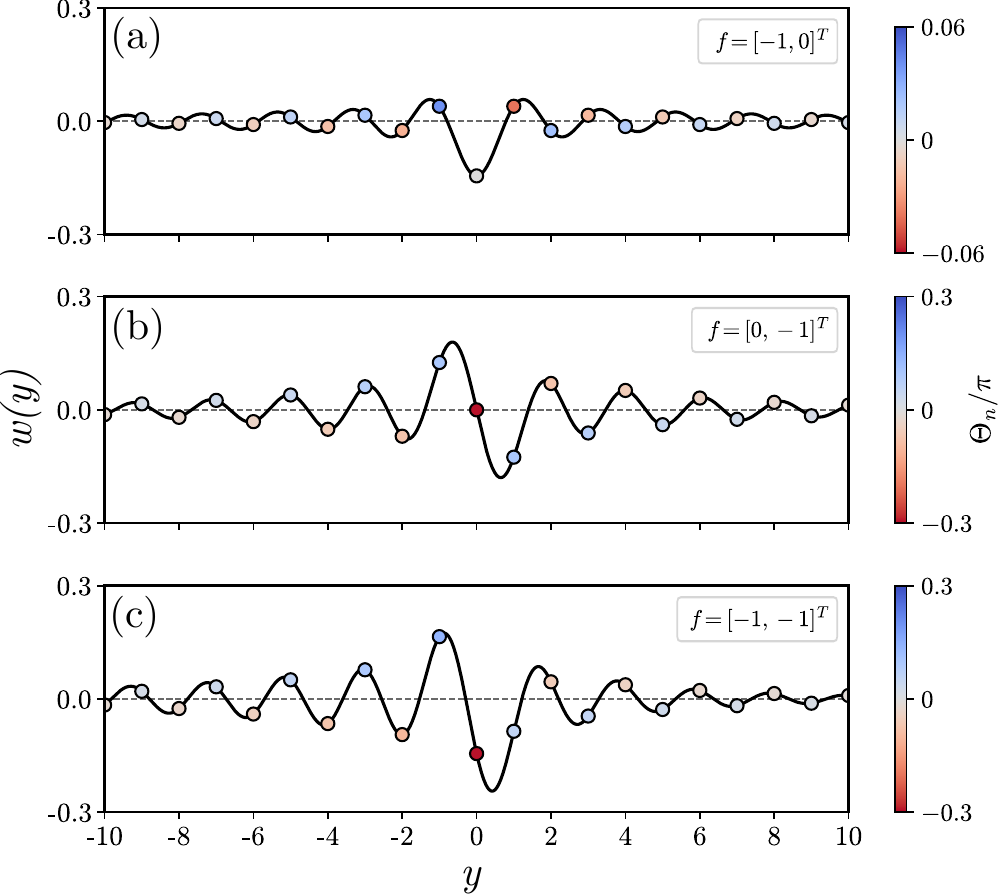}
    \caption{Green's function solution in the band gap for the $p = 3$ lattice as in Fig~\ref{Fig: 1D Dispersion}. Excitation frequency is $\omega = 8$ for three cases of generalised forcing vector, $\mathbf{f}$: (a) and (b) show purely translation and rotational forcing, respectively. (c) shows a mixed forcing.}
    \label{Fig: GreensFunction}
\end{figure}

However, our interpretation of the band edge solutions in this way is only valid if
\begin{align}
    \begin{split}
        \mu < 
        \frac{\left(1+ \sum\limits_{p>1, \text{odd}} p^{-1}+\sum\limits_{p>1, \text{even}} 3p^{-1} \right)}{12\left(1  +\sum\limits_{p>1, \text{odd}} p^{-1}\right)},
        \label{Eq: Band gap}
    \end{split}
\end{align}
for $p\in\mathcal{P}$.
The band gap closes when $\mu$ is equal to the right hand side of \eqref{Eq: Band gap}. Subsequently, if $\mu$ is greater than \eqref{Eq: Band gap}, the solutions re-order such that, $\omega _1^2 \leftrightarrow\omega _2^2$. The interchangeability of the band edge solutions based solely on varying the rotational inertia emphasises the pivotal role $\mu$ has on the overall lattice behaviour, the impact of which is not limited to the width and position of the band gap \cite{Madine2021}.

We gain further insight into the dynamics of the system by extending our analysis to the inhomogenous, forced problem,  allowing us to extract decaying solutions in the band gaps. In Fig.~\ref{Fig: GreensFunction}, we explore solutions in the band gap for the case of the $p=3$ lattice by numerically evaluating the dynamic Green's function for three different applied forcings; the two degrees-of-freedom in the flexural lattice enables both translational displacements and rotational moments to be induced. Through introducion of a generalised applied force to the central mass,
$\boldsymbol{f} = 
 \begin{bmatrix}
            f_w & f_{\theta}\\
\end{bmatrix}^T$,   
where $\boldsymbol{f} \in \mathbb{C}^2$, we construct the Green’s functions through (17). Here $f_w$ is the applied point translational force, and $f_{\theta}$ is the rotational point moment. 

The inclusion of the applied forcing vector permits the re-expression of the equation of motion to yield the Green’s Function in reciprocal space:
\begin{eqnarray}
    \boldsymbol{U}=G^{-1}\boldsymbol{f}.
\end{eqnarray}
Subsequently, application of the inverse Fourier transform (which we do numerically) results in the real space Green's Function, shown in Figure~\ref{Fig: GreensFunction}(a) for an applied translational force, with Fig.~\ref{Fig: GreensFunction}(b) demonstrating an applied rotational moment, and finally Figure \ref{Fig: GreensFunction}(c) depicting a combined translational and rotational forcing. In each case, the driving frequency is $\omega = 8$, within the band gap as can bee seen in Fig.~\ref{Fig: 1D Dispersion}(a).

The combined effect of both rotational and translational applied forcing, results in asymmetric wave excitation \cite{KatieThesis}. Due to the handedness of the moment applied, the right-hand side of the central node experiences a more rapidly decaying envelope as a result of the rotational and translational modes being out of phase. The general form of the decay envelope as a function of frequency can be predicted from, for example, high frequency homogenisation \cite{craster2010high}.

Until now we have only considered the band gap behaviour and the labelling of the branches due to the flavour of the standing wave solutions at the band edge. The inclusion of BNN couplings within the mass-beam lattices induces points of extrema within the first BZ, a behaviour comparable to that observed in mass-spring systems where these extrema are dependent on the relative strength and range of the BNN interactions \cite{Chen2021}. We note here that we only tailor the spatial range of the BNN connections and not the coupling strengths; the beams material parameters $(E,I, A,\rho)$ are normalised and identical across NN and BNN alike. Moreover, contrary to the mass-spring model, each branch experiences different susceptibilities to the extrema due to the order of the governing equation. Figure~\ref{Fig: 1D Dispersion}(d) highlights the unusual mode shapes associated with the effective beam profile at regions of ZGV within the first BZ. The mode shape clearly does not represent a conventional sinusoidal standing wave, which leads us to evaluate the energy flux associated with each displacement component. 

\begin{figure}
    \centering
    \includegraphics[width=0.375\textwidth]{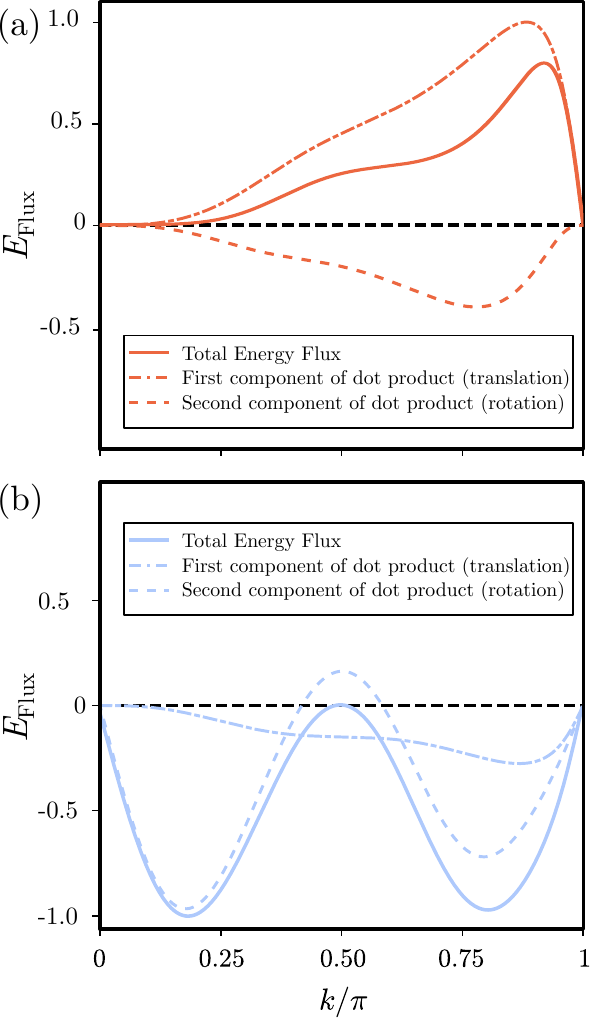}
    \caption{Normalised Energy Flux. (a) Total normalised Energy flux (solid line) for the lower branch of $p = 3$ BNN dispersion curve. Dashed-dotted and dashed lines show the first and second components of the dot product in \eqref{eq:flux} respectively, corresponding to translational and rotational `components'. (b) analogous curves but for the upper branch.}
    \label{fig:1D Energy Flux}
\end{figure}

The existence of multiple deformation pathways, combined with BNN lattice connections, gives rise to competing power channels that have been claimed to induce points of ZGV \cite{Chaplainreconfig}. We qualify this here in Figure~\ref{fig:1D Energy Flux}, showing the rate of energy transfer between unit cells, incorporating the stiffness matrices and the generalised forcing vectors in order to characterise both the translational and rotational degrees-of-freedom, leading to
\begin{align}
    \begin{split}
        E_{\mathrm{Flux}} = \Re \left[\left(i\omega \boldsymbol{U_n}\right)^{\dagger}\cdot\left( \sum_{p\in \mathcal{P}}\ p\left(B_p^+ e^{ikp}-A_p^+ \right)\right)\boldsymbol{U_n} \right].
        \label{eq:flux}
    \end{split}
\end{align}

\begin{figure}
    \centering
    \includegraphics[width=0.4\textwidth]{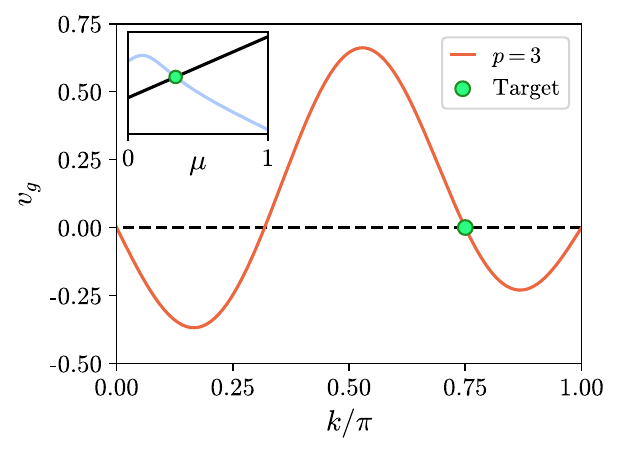}
    \caption{Example tuning of $k$ value of ZGV. Here we tune a ZGV mode to occur at $k=0.75\pi$ for the $p=3$ lattice by tailoring the rotational inertial of $\mu=0.34$. Inset shows the graphically solution of \eqref{Eq: Tuned Vg=0 Eq}.}
    \label{fig:Tuned Vg}
\end{figure}

Following the recent interest surrounding Roton-like dispersion \cite{Kazemi2023}, we plot the resultant energy flux along the two dispersion branches of the $p=3$ BNN lattice in Figure~\ref{fig:1D Energy Flux}. The total energy flux in Fig.~\ref{fig:1D Energy Flux}(b) shows that at $k/\pi = 0.5$ there is vanishing energy propagation along the array, aligning with the ZGV mode in Fig.~\ref{Fig: 1D Dispersion}(d), despite it not representing a conventional standing wave, as discussed. We confirm through numerical calculation that the total energy flux associated with the mode at this wavenumber is zero and as such there is no power transfer along the lattice.  We also show the 
`components' of the energy flux calculation associated with the individual degrees-of-freedom, i.e. the first and second terms elements the dot product in \eqref{eq:flux}, highlighting the competing nature of the power channels underpinning the extrema in the dispersion curves.

Having investigated a mechanism for driving points of ZGV in BNN connected discrete lattices, we employ a simple method to tune the positions in reciprocal space by altering the rotational inertia of the lattice sites, and provide motivation in the context of metamaterial energy harvesting devices.

\subsection*{\label{Sec: Tuning ZGV}Tuning Points of ZGV}

ZGV modes have been shown to exhibit enhanced energy harvesting in metamaterial devices combined with grading, owing to the slowing of waves on approach to points of local ZGV, thereby increasing the interaction time with harvesting devices \cite{de2020graded,Chaplainrainbowreflection,DePontiHarvesting}. For mass-spring systems with BNN couplings (and unit cell size $a$), positions of extrema within the first BZ occur at intervals of $\frac{\pi}{pa}$ due to extraordinary Bragg reflections \cite{Chen2021}, provided the coupling strength is commensurate with NN (i.e. a competing power channel). Conversely, for mass-beam lattices, which possess coupled interactions between rotational and translational behaviours, the location of the dispersion extrema is not as established. Consequently, we explore the tuning of ZGV modes in reciprocal space to predetermined $k$ values by manipulating the second degree-of-freedom through the rotational inertia and hence focus on the second branch.

The targeted tailoring of the rotational inertia, to tune a point of ZGV, requires \eqref{Eq: Vg=0} to be solved for $\mu$. We arbitrarily define the BNN connection orders $p = 3$ and the intended location of the tuned ZGV mode $(k)$ to $k=0.75\pi$. Additionally, we re-express the analytical ZGV 
 equation, \eqref{Eq: Vg=0} as
\begin{align}
    \begin{split}
        \frac{d}{dk} \left( 24\mu \alpha +\beta \right) = \mp \frac{d}{dk} \left(\left(24\mu \alpha-\beta \right)^2-4\mu \gamma ^2 \right)^\frac{1}{2},
    \end{split}
    \label{Eq: Tuned Vg=0 Eq}
\end{align}
where the left and right-hand side of the equation are polynomials in terms of $\mu$, which we solve for numerically; determining the value of $\mu$ at which the polynomials intersect. Any zero solution obtained is treated as nonphysical. This is shown graphically by the insert in Figure~\ref{fig:Tuned Vg} which intersects at $\mu=0.34$. We then validate by calculating the tuned group velocity for the obtained $\mu$ value, illustrated in Figure \ref{fig:Tuned Vg}. As expected, the lattice maintains ZGV modes at the band edge and centre, however a point of ZGV has been tuned to $k=0.75\pi$ as intended, satisfying the design paradigm.

\section{\label{Sec: Conclusions}Conclusions}

Elastic waves that propagate on discrete mass-beam lattices with BNN connections have unique dispersive properties that arise through the coupling between translational and rotational degrees-of-freedom at the nodal junctions, and by spatial connections that change the locality of dispersion extrema. By considering the dynamics of EB lattices at nodal positions, akin to discrete mass-spring systems, analytical and numerical analyses were employed to demonstrate flexibility in studying and tuning the dynamics of such lattices. 

We provided motivation for the mechanism that drives extrema in the dispersion curves within the first BZ, namely the introduction of competing power channels, that we verified by considering the power flow across a unit cell. 

Dispersion engineering and design has garnered recent attention with the use of BNNs, and in particular the tuning of ZGV modes has promising applications in energy harvesting devices. The simplistic models presented offer insight and intuition into the governing dynamics of more complex systems, both elastic systems, and in other wave regimes. We provide numerical evaluation of both the eigensolutions to the homogeneous problem, and the dynamical Green's function to the forced problem, providing a complete toolkit to visualise complex wave phenomenon in flexural systems with BNN connections.

The realisation of physical systems with many BNN connections, with potentially varying material parameters, may be ill-suited to passive structures and hence motivates the use of alternative methods such as active control. We foresee dynamic feedback mechanisms as a promising route to provide an experimental verification, for potentially arbitrary many virtual BNN connections (that is inherently re-configurable), mimicking the lattice deformations and the interactions of higher order beam connections in an analogous fashion to the effective beam displacements we show throughout.

\begin{acknowledgements} 
\noindent R.G.E. and T.A.S. acknowledge the financial support of Defence Science and Technology Laboratory (Dstl) through grants, DSTL0000022047, DSTLXR1000154754 and AGR 0117701. E.P and G.J.C acknowledge the financial support by the EPSRC (grant no EP/Y015673/1). K.H.M. and D.J.C. gratefully acknowledge funding from The Leverhulme Trust through Research Project Grant RPG-2022-261. G.J.C. also gratefully acknowledges financial support from the Royal Commission for the Exhibition of 1851 in the form of a Research Fellowship. All data created during this research are available upon reasonable request to the corresponding author. ‘For the purpose of open access, the author has applied a ‘Creative Commons Attribution (CC BY) licence to any Author Accepted Manuscript version arising from this submission’.
\end{acknowledgements}


\begin{thebibliography}{23}%
\makeatletter
\providecommand \@ifxundefined [1]{%
 \@ifx{#1\undefined}
}%
\providecommand \@ifnum [1]{%
 \ifnum #1\expandafter \@firstoftwo
 \else \expandafter \@secondoftwo
 \fi
}%
\providecommand \@ifx [1]{%
 \ifx #1\expandafter \@firstoftwo
 \else \expandafter \@secondoftwo
 \fi
}%
\providecommand \natexlab [1]{#1}%
\providecommand \enquote  [1]{``#1''}%
\providecommand \bibnamefont  [1]{#1}%
\providecommand \bibfnamefont [1]{#1}%
\providecommand \citenamefont [1]{#1}%
\providecommand \href@noop [0]{\@secondoftwo}%
\providecommand \href [0]{\begingroup \@sanitize@url \@href}%
\providecommand \@href[1]{\@@startlink{#1}\@@href}%
\providecommand \@@href[1]{\endgroup#1\@@endlink}%
\providecommand \@sanitize@url [0]{\catcode `\\12\catcode `\$12\catcode `\&12\catcode `\#12\catcode `\^12\catcode `\_12\catcode `\%12\relax}%
\providecommand \@@startlink[1]{}%
\providecommand \@@endlink[0]{}%
\providecommand \url  [0]{\begingroup\@sanitize@url \@url }%
\providecommand \@url [1]{\endgroup\@href {#1}{\urlprefix }}%
\providecommand \urlprefix  [0]{URL }%
\providecommand \Eprint [0]{\href }%
\providecommand \doibase [0]{https://doi.org/}%
\providecommand \selectlanguage [0]{\@gobble}%
\providecommand \bibinfo  [0]{\@secondoftwo}%
\providecommand \bibfield  [0]{\@secondoftwo}%
\providecommand \translation [1]{[#1]}%
\providecommand \BibitemOpen [0]{}%
\providecommand \bibitemStop [0]{}%
\providecommand \bibitemNoStop [0]{.\EOS\space}%
\providecommand \EOS [0]{\spacefactor3000\relax}%
\providecommand \BibitemShut  [1]{\csname bibitem#1\endcsname}%
\let\auto@bib@innerbib\@empty
\bibitem [{\citenamefont {Brillouin}(1953)}]{Brillouin1953}%
  \BibitemOpen
  \bibfield  {author} {\bibinfo {author} {\bibfnamefont {L.}~\bibnamefont {Brillouin}},\ }\href@noop {} {\emph {\bibinfo {title} {Wave Propagation in Periodic Structures: Electric Filters and Crystal Lattices}}},\ \bibinfo {edition} {2nd}\ ed.\ (\bibinfo  {publisher} {Dover Publications},\ \bibinfo {year} {1953})\BibitemShut {NoStop}%
\bibitem [{\citenamefont {Chen}\ \emph {et~al.}(2021)\citenamefont {Chen}, \citenamefont {Kadic},\ and\ \citenamefont {Wegener}}]{Chen2021}%
  \BibitemOpen
  \bibfield  {author} {\bibinfo {author} {\bibfnamefont {Y.}~\bibnamefont {Chen}}, \bibinfo {author} {\bibfnamefont {M.}~\bibnamefont {Kadic}},\ and\ \bibinfo {author} {\bibfnamefont {M.}~\bibnamefont {Wegener}},\ }\bibfield  {title} {\bibinfo {title} {Roton-like acoustical dispersion relations in 3d metamaterials},\ }\href {https://doi.org/10.1038/s41467-021-23574-2} {\bibfield  {journal} {\bibinfo  {journal} {Nature Communications 2021 12:1}\ }\textbf {\bibinfo {volume} {12}},\ \bibinfo {pages} {1} (\bibinfo {year} {2021})}\BibitemShut {NoStop}%
\bibitem [{\citenamefont {Madine}\ and\ \citenamefont {Colquitt}(2021)}]{Madine2021}%
  \BibitemOpen
  \bibfield  {author} {\bibinfo {author} {\bibfnamefont {K.~H.}\ \bibnamefont {Madine}}\ and\ \bibinfo {author} {\bibfnamefont {D.~J.}\ \bibnamefont {Colquitt}},\ }\bibfield  {title} {\bibinfo {title} {Dynamic green’s functions in discrete flexural systems},\ }\href {https://doi.org/10.1093/QJMAM/HBAB006} {\bibfield  {journal} {\bibinfo  {journal} {The Quarterly Journal of Mechanics and Applied Mathematics}\ }\textbf {\bibinfo {volume} {74}},\ \bibinfo {pages} {323} (\bibinfo {year} {2021})}\BibitemShut {NoStop}%
\bibitem [{\citenamefont {Wang}(2013)}]{Wang2013}%
  \BibitemOpen
  \bibfield  {author} {\bibinfo {author} {\bibfnamefont {G.}~\bibnamefont {Wang}},\ }\bibfield  {title} {\bibinfo {title} {Analysis of bimorph piezoelectric beam energy harvesters using timoshenko and euler-bernoulli beam theory},\ }\href {https://doi.org/10.1177/1045389X12461080/ASSET/IMAGES/LARGE/10.1177_1045389X12461080-FIG10.JPEG} {\bibfield  {journal} {\bibinfo  {journal} {Journal of Intelligent Material Systems and Structures}\ }\textbf {\bibinfo {volume} {24}},\ \bibinfo {pages} {226} (\bibinfo {year} {2013})}\BibitemShut {NoStop}%
\bibitem [{\citenamefont {Caddemi}\ and\ \citenamefont {Morassi}(2012)}]{Caddemi2012}%
  \BibitemOpen
  \bibfield  {author} {\bibinfo {author} {\bibfnamefont {S.}~\bibnamefont {Caddemi}}\ and\ \bibinfo {author} {\bibfnamefont {A.}~\bibnamefont {Morassi}},\ }\bibfield  {title} {\bibinfo {title} {Multi-cracked euler-bernoulli beams: Mathematical modeling and exact solutions},\ }\bibfield  {journal} {\bibinfo  {journal} {International Journal of Solids and Structures}\ }\href {https://doi.org/10.1016/j.ijsolstr.2012.11.018} {10.1016/j.ijsolstr.2012.11.018} (\bibinfo {year} {2012})\BibitemShut {NoStop}%
\bibitem [{\citenamefont {Fernández-Sáez}\ \emph {et~al.}(2015)\citenamefont {Fernández-Sáez}, \citenamefont {Zaera}, \citenamefont {Loya},\ and\ \citenamefont {Reddy}}]{Eringen}%
  \BibitemOpen
  \bibfield  {author} {\bibinfo {author} {\bibfnamefont {J.}~\bibnamefont {Fernández-Sáez}}, \bibinfo {author} {\bibfnamefont {R.}~\bibnamefont {Zaera}}, \bibinfo {author} {\bibfnamefont {J.~A.}\ \bibnamefont {Loya}},\ and\ \bibinfo {author} {\bibfnamefont {J.~N.}\ \bibnamefont {Reddy}},\ }\bibfield  {title} {\bibinfo {title} {Bending of euler-bernoulli beams using eringen's integral formulation: A paradox resolved},\ }\href {https://doi.org/10.1016/j.ijengsci.2015.10.013} {\bibfield  {journal} {\bibinfo  {journal} {International Journal of Engineering Science}\ }\textbf {\bibinfo {volume} {99}},\ \bibinfo {pages} {107} (\bibinfo {year} {2015})}\BibitemShut {NoStop}%
\bibitem [{\citenamefont {Madine}\ and\ \citenamefont {Colquitt}(2022)}]{Madine2022}%
  \BibitemOpen
  \bibfield  {author} {\bibinfo {author} {\bibfnamefont {K.~H.}\ \bibnamefont {Madine}}\ and\ \bibinfo {author} {\bibfnamefont {D.~J.}\ \bibnamefont {Colquitt}},\ }\bibfield  {title} {\bibinfo {title} {Negative refraction and mode trapping of flexural–torsional waves in elastic lattices},\ }\bibfield  {journal} {\bibinfo  {journal} {Philosophical Transactions of the Royal Society A}\ }\textbf {\bibinfo {volume} {380}},\ \href {https://doi.org/10.1098/RSTA.2021.0379} {10.1098/RSTA.2021.0379} (\bibinfo {year} {2022})\BibitemShut {NoStop}%
\bibitem [{\citenamefont {Chaplain}\ \emph {et~al.}(2023)\citenamefont {Chaplain}, \citenamefont {Hooper}, \citenamefont {Hibbins},\ and\ \citenamefont {Starkey}}]{Chaplainreconfig}%
  \BibitemOpen
  \bibfield  {author} {\bibinfo {author} {\bibfnamefont {G.~J.}\ \bibnamefont {Chaplain}}, \bibinfo {author} {\bibfnamefont {I.~R.}\ \bibnamefont {Hooper}}, \bibinfo {author} {\bibfnamefont {A.~P.}\ \bibnamefont {Hibbins}},\ and\ \bibinfo {author} {\bibfnamefont {T.~A.}\ \bibnamefont {Starkey}},\ }\bibfield  {title} {\bibinfo {title} {Reconfigurable elastic metamaterials: Engineering dispersion with beyond nearest neighbors},\ }\href {https://doi.org/10.1103/PHYSREVAPPLIED.19.044061/FIGURES/4/MEDIUM} {\bibfield  {journal} {\bibinfo  {journal} {Physical Review Applied}\ }\textbf {\bibinfo {volume} {19}},\ \bibinfo {pages} {044061} (\bibinfo {year} {2023})}\BibitemShut {NoStop}%
\bibitem [{\citenamefont {Kazemi}\ \emph {et~al.}(2023)\citenamefont {Kazemi}, \citenamefont {Deshmukh}, \citenamefont {Chen}, \citenamefont {Liu}, \citenamefont {Deng}, \citenamefont {Fu},\ and\ \citenamefont {Wang}}]{Kazemi2023}%
  \BibitemOpen
  \bibfield  {author} {\bibinfo {author} {\bibfnamefont {A.}~\bibnamefont {Kazemi}}, \bibinfo {author} {\bibfnamefont {K.~J.}\ \bibnamefont {Deshmukh}}, \bibinfo {author} {\bibfnamefont {F.}~\bibnamefont {Chen}}, \bibinfo {author} {\bibfnamefont {Y.}~\bibnamefont {Liu}}, \bibinfo {author} {\bibfnamefont {B.}~\bibnamefont {Deng}}, \bibinfo {author} {\bibfnamefont {H.~C.}\ \bibnamefont {Fu}},\ and\ \bibinfo {author} {\bibfnamefont {P.}~\bibnamefont {Wang}},\ }\bibfield  {title} {\bibinfo {title} {Drawing dispersion curves: Band structure customization via nonlocal phononic crystals},\ }\href {https://doi.org/10.1103/PhysRevLett.131.176101} {\bibfield  {journal} {\bibinfo  {journal} {Physical Review Letters}\ }\textbf {\bibinfo {volume} {131}},\ \bibinfo {pages} {176101} (\bibinfo {year} {2023})}\BibitemShut {NoStop}%
\bibitem [{\citenamefont {Pu}\ \emph {et~al.}(2010)\citenamefont {Pu}, \citenamefont {Dong}, \citenamefont {Huang}, \citenamefont {Barough}, \citenamefont {Noori}, \citenamefont {Abbasiyan}, \citenamefont {Schulz}, \citenamefont {O'faolain}, \citenamefont {Beggs}, \citenamefont {White}, \citenamefont {Melloni},\ and\ \citenamefont {Krauss}}]{Pu2010}%
  \BibitemOpen
  \bibfield  {author} {\bibinfo {author} {\bibfnamefont {S.}~\bibnamefont {Pu}}, \bibinfo {author} {\bibfnamefont {S.}~\bibnamefont {Dong}}, \bibinfo {author} {\bibfnamefont {J.}~\bibnamefont {Huang}}, \bibinfo {author} {\bibfnamefont {A.~S.}\ \bibnamefont {Barough}}, \bibinfo {author} {\bibfnamefont {M.}~\bibnamefont {Noori}}, \bibinfo {author} {\bibfnamefont {A.}~\bibnamefont {Abbasiyan}}, \bibinfo {author} {\bibfnamefont {S.~A.}\ \bibnamefont {Schulz}}, \bibinfo {author} {\bibfnamefont {L.}~\bibnamefont {O'faolain}}, \bibinfo {author} {\bibfnamefont {D.~M.}\ \bibnamefont {Beggs}}, \bibinfo {author} {\bibfnamefont {T.~P.}\ \bibnamefont {White}}, \bibinfo {author} {\bibfnamefont {A.}~\bibnamefont {Melloni}},\ and\ \bibinfo {author} {\bibfnamefont {T.~F.}\ \bibnamefont {Krauss}},\ }\bibfield  {title} {\bibinfo {title} {Dispersion engineered slow light in photonic crystals: a comparison},\ }\href {https://doi.org/10.1088/2040-8978/12/10/104004} {\bibfield  {journal} {\bibinfo  {journal} {Journal of Optics}\
  }\textbf {\bibinfo {volume} {12}},\ \bibinfo {pages} {104004} (\bibinfo {year} {2010})}\BibitemShut {NoStop}%
\bibitem [{\citenamefont {Smith}\ \emph {et~al.}(2004)\citenamefont {Smith}, \citenamefont {Pendry},\ and\ \citenamefont {Wiltshire}}]{Smith2004}%
  \BibitemOpen
  \bibfield  {author} {\bibinfo {author} {\bibfnamefont {D.~R.}\ \bibnamefont {Smith}}, \bibinfo {author} {\bibfnamefont {J.~B.}\ \bibnamefont {Pendry}},\ and\ \bibinfo {author} {\bibfnamefont {M.~C.}\ \bibnamefont {Wiltshire}},\ }\bibfield  {title} {\bibinfo {title} {Metamaterials and negative refractive index},\ }\href {https://doi.org/10.1126/SCIENCE.1096796/ASSET/AE0DF28E-3FA8-42B0-9269-AFAD6619A0AD/ASSETS/GRAPHIC/ZSE0310427410004.JPEG} {\bibfield  {journal} {\bibinfo  {journal} {Science}\ }\textbf {\bibinfo {volume} {305}},\ \bibinfo {pages} {788} (\bibinfo {year} {2004})}\BibitemShut {NoStop}%
\bibitem [{\citenamefont {Craster}\ and\ \citenamefont {Guenneau}(2013)}]{Craster2013}%
  \BibitemOpen
  \bibfield  {author} {\bibinfo {author} {\bibfnamefont {R.}~\bibnamefont {Craster}}\ and\ \bibinfo {author} {\bibfnamefont {S.}~\bibnamefont {Guenneau}},\ }\href {https://doi.org/10.1007/978-94-007-4813-2} {\emph {\bibinfo {title} {Acoustic Metamaterials: Negative Refraction, Imaging, Lensing and Cloaking}}},\ edited by\ \bibinfo {editor} {\bibfnamefont {R.~V.}\ \bibnamefont {Craster}}\ and\ \bibinfo {editor} {\bibfnamefont {S.}~\bibnamefont {Guenneau}},\ Vol.\ \bibinfo {volume} {166}\ (\bibinfo  {publisher} {Springer Netherlands},\ \bibinfo {year} {2013})\BibitemShut {NoStop}%
\bibitem [{\citenamefont {Chaplain}\ \emph {et~al.}(2020{\natexlab{a}})\citenamefont {Chaplain}, \citenamefont {Pajer}, \citenamefont {Ponti},\ and\ \citenamefont {Craster}}]{Chaplainrainbowreflection}%
  \BibitemOpen
  \bibfield  {author} {\bibinfo {author} {\bibfnamefont {G.~J.}\ \bibnamefont {Chaplain}}, \bibinfo {author} {\bibfnamefont {D.}~\bibnamefont {Pajer}}, \bibinfo {author} {\bibfnamefont {J.~M.~D.}\ \bibnamefont {Ponti}},\ and\ \bibinfo {author} {\bibfnamefont {R.~V.}\ \bibnamefont {Craster}},\ }\bibfield  {title} {\bibinfo {title} {Delineating rainbow reflection and trapping with applications for energy harvesting},\ }\bibfield  {journal} {\bibinfo  {journal} {New Journal of Physics}\ }\textbf {\bibinfo {volume} {22}},\ \href {https://doi.org/10.1088/1367-2630/AB8CAE} {10.1088/1367-2630/AB8CAE} (\bibinfo {year} {2020}{\natexlab{a}})\BibitemShut {NoStop}%
\bibitem [{\citenamefont {Chaplain}\ \emph {et~al.}(2020{\natexlab{b}})\citenamefont {Chaplain}, \citenamefont {Ponti}, \citenamefont {Aguzzi}, \citenamefont {Colombi},\ and\ \citenamefont {Craster}}]{Chaplainrainbowtrapping}%
  \BibitemOpen
  \bibfield  {author} {\bibinfo {author} {\bibfnamefont {G.~J.}\ \bibnamefont {Chaplain}}, \bibinfo {author} {\bibfnamefont {J.~M.~D.}\ \bibnamefont {Ponti}}, \bibinfo {author} {\bibfnamefont {G.}~\bibnamefont {Aguzzi}}, \bibinfo {author} {\bibfnamefont {A.}~\bibnamefont {Colombi}},\ and\ \bibinfo {author} {\bibfnamefont {R.~V.}\ \bibnamefont {Craster}},\ }\bibfield  {title} {\bibinfo {title} {Topological rainbow trapping for elastic energy harvesting in graded su-schrieffer-heeger systems},\ }\href {https://doi.org/10.1103/PHYSREVAPPLIED.14.054035/FIGURES/10/MEDIUM} {\bibfield  {journal} {\bibinfo  {journal} {Physical Review Applied}\ }\textbf {\bibinfo {volume} {14}},\ \bibinfo {pages} {054035} (\bibinfo {year} {2020}{\natexlab{b}})}\BibitemShut {NoStop}%
\bibitem [{\citenamefont {Nassar}\ \emph {et~al.}(2020)\citenamefont {Nassar}, \citenamefont {Yousefzadeh}, \citenamefont {Fleury}, \citenamefont {Ruzzene}, \citenamefont {Alù}, \citenamefont {Daraio}, \citenamefont {Norris}, \citenamefont {Huang},\ and\ \citenamefont {Haberman}}]{Nassar2020}%
  \BibitemOpen
  \bibfield  {author} {\bibinfo {author} {\bibfnamefont {H.}~\bibnamefont {Nassar}}, \bibinfo {author} {\bibfnamefont {B.}~\bibnamefont {Yousefzadeh}}, \bibinfo {author} {\bibfnamefont {R.}~\bibnamefont {Fleury}}, \bibinfo {author} {\bibfnamefont {M.}~\bibnamefont {Ruzzene}}, \bibinfo {author} {\bibfnamefont {A.}~\bibnamefont {Alù}}, \bibinfo {author} {\bibfnamefont {C.}~\bibnamefont {Daraio}}, \bibinfo {author} {\bibfnamefont {A.~N.}\ \bibnamefont {Norris}}, \bibinfo {author} {\bibfnamefont {G.}~\bibnamefont {Huang}},\ and\ \bibinfo {author} {\bibfnamefont {M.~R.}\ \bibnamefont {Haberman}},\ }\bibfield  {title} {\bibinfo {title} {Nonreciprocity in acoustic and elastic materials},\ }\href {https://doi.org/10.1038/S41578-020-0206-0} {\bibfield  {journal} {\bibinfo  {journal} {Nature Reviews Materials}\ }\textbf {\bibinfo {volume} {5}},\ \bibinfo {pages} {667} (\bibinfo {year} {2020})}\BibitemShut {NoStop}%
\bibitem [{\citenamefont {Frenzel}\ \emph {et~al.}(2017)\citenamefont {Frenzel}, \citenamefont {Kadic},\ and\ \citenamefont {Wegener}}]{Frenzel2017}%
  \BibitemOpen
  \bibfield  {author} {\bibinfo {author} {\bibfnamefont {T.}~\bibnamefont {Frenzel}}, \bibinfo {author} {\bibfnamefont {M.}~\bibnamefont {Kadic}},\ and\ \bibinfo {author} {\bibfnamefont {M.}~\bibnamefont {Wegener}},\ }\bibfield  {title} {\bibinfo {title} {Three-dimensional mechanical metamaterials with a twist},\ }\href {https://doi.org/10.1126/SCIENCE.AAO4640/SUPPL_FILE/AAO4640_MOVIE_S1.MP4} {\bibfield  {journal} {\bibinfo  {journal} {Science}\ }\textbf {\bibinfo {volume} {358}},\ \bibinfo {pages} {1072} (\bibinfo {year} {2017})}\BibitemShut {NoStop}%
\bibitem [{\citenamefont {Martínez}\ \emph {et~al.}(2021)\citenamefont {Martínez}, \citenamefont {Groß}, \citenamefont {Chen}, \citenamefont {Frenzel}, \citenamefont {Laude}, \citenamefont {Kadic},\ and\ \citenamefont {Wegener}}]{Martinez}%
  \BibitemOpen
  \bibfield  {author} {\bibinfo {author} {\bibfnamefont {J.~A.~I.}\ \bibnamefont {Martínez}}, \bibinfo {author} {\bibfnamefont {M.~F.}\ \bibnamefont {Groß}}, \bibinfo {author} {\bibfnamefont {Y.}~\bibnamefont {Chen}}, \bibinfo {author} {\bibfnamefont {T.}~\bibnamefont {Frenzel}}, \bibinfo {author} {\bibfnamefont {V.}~\bibnamefont {Laude}}, \bibinfo {author} {\bibfnamefont {M.}~\bibnamefont {Kadic}},\ and\ \bibinfo {author} {\bibfnamefont {M.}~\bibnamefont {Wegener}},\ }\bibfield  {title} {\bibinfo {title} {Experimental observation of roton-like dispersion relations in metamaterials},\ }\href {https://doi.org/10.1126/SCIADV.ABM2189/SUPPL_FILE/SCIADV.ABM2189_SM.PDF} {\bibfield  {journal} {\bibinfo  {journal} {Science Advances}\ }\textbf {\bibinfo {volume} {7}},\ \bibinfo {pages} {2189} (\bibinfo {year} {2021})}\BibitemShut {NoStop}%
\bibitem [{\citenamefont {Graff}(1991)}]{Graff}%
  \BibitemOpen
  \bibfield  {author} {\bibinfo {author} {\bibfnamefont {K.~F.}\ \bibnamefont {Graff}},\ }\href@noop {} {\emph {\bibinfo {title} {Wave Motion in Elastic Solids}}},\ \bibinfo {edition} {reprint}\ ed.\ (\bibinfo  {publisher} {Dover Publucations},\ \bibinfo {year} {1991})\BibitemShut {NoStop}%
\bibitem [{\citenamefont {Jiao}\ \emph {et~al.}(2019)\citenamefont {Jiao}, \citenamefont {Alavi}, \citenamefont {Borchani},\ and\ \citenamefont {Lajnef}}]{Jiao2019}%
  \BibitemOpen
  \bibfield  {author} {\bibinfo {author} {\bibfnamefont {P.}~\bibnamefont {Jiao}}, \bibinfo {author} {\bibfnamefont {A.~H.}\ \bibnamefont {Alavi}}, \bibinfo {author} {\bibfnamefont {W.}~\bibnamefont {Borchani}},\ and\ \bibinfo {author} {\bibfnamefont {N.}~\bibnamefont {Lajnef}},\ }\bibfield  {title} {\bibinfo {title} {Small and large deformation models of post-buckled beams under lateral constraints},\ }\href {https://doi.org/10.1177/1081286517741341/ASSET/IMAGES/LARGE/10.1177_1081286517741341-FIG14.JPEG} {\bibfield  {journal} {\bibinfo  {journal} {Mathematics and Mechanics of Solids}\ }\textbf {\bibinfo {volume} {24}},\ \bibinfo {pages} {386} (\bibinfo {year} {2019})}\BibitemShut {NoStop}%
\bibitem [{\citenamefont {Madine}(2023)}]{KatieThesis}%
  \BibitemOpen
  \bibfield  {author} {\bibinfo {author} {\bibfnamefont {K.}~\bibnamefont {Madine}},\ }\emph {\bibinfo {title} {Wave Propagation in Flexural Systems}},\ \href@noop {} {Ph.D. thesis},\ \bibinfo  {school} {University of Liverpool} (\bibinfo {year} {2023})\BibitemShut {NoStop}%
\bibitem [{\citenamefont {Craster}\ \emph {et~al.}(2010)\citenamefont {Craster}, \citenamefont {Kaplunov},\ and\ \citenamefont {Pichugin}}]{craster2010high}%
  \BibitemOpen
  \bibfield  {author} {\bibinfo {author} {\bibfnamefont {R.~V.}\ \bibnamefont {Craster}}, \bibinfo {author} {\bibfnamefont {J.}~\bibnamefont {Kaplunov}},\ and\ \bibinfo {author} {\bibfnamefont {A.~V.}\ \bibnamefont {Pichugin}},\ }\bibfield  {title} {\bibinfo {title} {High-frequency homogenization for periodic media},\ }\href@noop {} {\bibfield  {journal} {\bibinfo  {journal} {Proceedings of the Royal Society A: Mathematical, Physical and Engineering Sciences}\ }\textbf {\bibinfo {volume} {466}},\ \bibinfo {pages} {2341} (\bibinfo {year} {2010})}\BibitemShut {NoStop}%
\bibitem [{\citenamefont {De~Ponti}\ \emph {et~al.}(2020)\citenamefont {De~Ponti}, \citenamefont {Colombi}, \citenamefont {Ardito}, \citenamefont {Braghin}, \citenamefont {Corigliano},\ and\ \citenamefont {Craster}}]{de2020graded}%
  \BibitemOpen
  \bibfield  {author} {\bibinfo {author} {\bibfnamefont {J.~M.}\ \bibnamefont {De~Ponti}}, \bibinfo {author} {\bibfnamefont {A.}~\bibnamefont {Colombi}}, \bibinfo {author} {\bibfnamefont {R.}~\bibnamefont {Ardito}}, \bibinfo {author} {\bibfnamefont {F.}~\bibnamefont {Braghin}}, \bibinfo {author} {\bibfnamefont {A.}~\bibnamefont {Corigliano}},\ and\ \bibinfo {author} {\bibfnamefont {R.~V.}\ \bibnamefont {Craster}},\ }\bibfield  {title} {\bibinfo {title} {Graded elastic metasurface for enhanced energy harvesting},\ }\href@noop {} {\bibfield  {journal} {\bibinfo  {journal} {New Journal of Physics}\ }\textbf {\bibinfo {volume} {22}},\ \bibinfo {pages} {013013} (\bibinfo {year} {2020})}\BibitemShut {NoStop}%
\bibitem [{\citenamefont {Ponti}\ \emph {et~al.}(2020)\citenamefont {Ponti}, \citenamefont {Colombi}, \citenamefont {Riva}, \citenamefont {Ardito}, \citenamefont {Braghin}, \citenamefont {Corigliano},\ and\ \citenamefont {Craster}}]{DePontiHarvesting}%
  \BibitemOpen
  \bibfield  {author} {\bibinfo {author} {\bibfnamefont {J.~M.~D.}\ \bibnamefont {Ponti}}, \bibinfo {author} {\bibfnamefont {A.}~\bibnamefont {Colombi}}, \bibinfo {author} {\bibfnamefont {E.}~\bibnamefont {Riva}}, \bibinfo {author} {\bibfnamefont {R.}~\bibnamefont {Ardito}}, \bibinfo {author} {\bibfnamefont {F.}~\bibnamefont {Braghin}}, \bibinfo {author} {\bibfnamefont {A.}~\bibnamefont {Corigliano}},\ and\ \bibinfo {author} {\bibfnamefont {R.~V.}\ \bibnamefont {Craster}},\ }\bibfield  {title} {\bibinfo {title} {Experimental investigation of amplification, via a mechanical delay-line, in a rainbow-based metamaterial for energy harvesting},\ }\href {https://doi.org/10.1063/5.0023544/1078485} {\bibfield  {journal} {\bibinfo  {journal} {Applied Physics Letters}\ }\textbf {\bibinfo {volume} {117}},\ \bibinfo {pages} {143902} (\bibinfo {year} {2020})}\BibitemShut {NoStop}%
\end{thebibliography}

%

\end{document}